\documentclass[sigconf]{acm-template/acmart}
\acmConference[SE 2030]{International Workshop on Software Engineering in 2030}{November 2024}{Puerto Galinàs (Brazil)}
\usepackage{graphicx} 
\usepackage{blindtext}
\usepackage{mathtools}
\usepackage{xspace}
\usepackage{tikz}

\newcommand{\subject}[0]{\emph{Subject}\xspace}

\newcommand{\ie}{i.e.,\xspace}
\newcommand{\eg}{e.g.,\xspace}
\newcommand{\etal}{et al.\xspace}

\title{A Roadmap for Simulation-Based Testing of Autonomous Cyber-Physical Systems: Challenges and Future Direction}

\author{Christian Birchler}
\email{christian.birchler@{zhaw,unibe}.ch}
\affiliation{
  \institution{Zurich University of Applied Sciences \\ University of Bern}
  \country{Switzerland}
}

\author{Sajad Khatiri}
\email{sajad.khatiri@zhaw.ch}
\affiliation{
  \institution{Zurich University of Applied Sciences}
  \country{Switzerland}
}

\author{Pooja Rani}
\email{rani@ifi.uzh.ch}
\affiliation{
  \institution{University of Zurich}
  \country{Switzerland}
}

\author{Timo Kehrer}
\email{timo.kehrer@unibe.ch}
\affiliation{
  \institution{University of Bern}
  \country{Switzerland}
}

\author{Sebastiano Panichella}
\email{sebastiano.panichella@zhaw.ch}
\affiliation{
  \institution{Zurich University of Applied Sciences}
  \country{Switzerland}
}

\begin{abstract}
As the era of autonomous cyber-physical systems (ACPSs), such as unmanned aerial vehicles and self-driving cars, unfolds, the demand for robust testing methodologies is key to realizing the adoption of such systems in real-world scenarios. However, traditional software testing paradigms face unprecedented challenges in ensuring the safety and reliability of these systems. In response, this paper pioneers a strategic roadmap for simulation-based testing of ACPSs, specifically focusing on autonomous systems.
Our paper discusses the relevant challenges and obstacles of ACPSs, focusing on test automation and quality assurance, hence advocating for tailored solutions to address the unique demands of autonomous systems. While providing concrete definitions of test cases within simulation environments, we also accentuate the need to create new benchmark assets and the development of automated tools tailored explicitly for autonomous systems in the software engineering community.
This paper not only highlights the relevant, pressing issues the software engineering community should focus on (in terms of practices, expected automation, and paradigms), but it also outlines ways to tackle them.
By outlining the various domains and challenges of simulation-based testing/development for ACPSs, we provide directions for future research efforts. 
\end{abstract}

\begin{document}

\maketitle

\section{Introduction}
In software engineering, testing stands as a cornerstone practice, essential for enhancing the reliability and robustness of software systems.
Test automation techniques, including test generators, selection strategies, and prioritization methods, are pivotal in reducing the need for costly manual, error-prone testing procedures.

With the contemporary rise of \emph{Autonomous Cyber-Physical Systems} (ACPSs), software engineers find themselves challenged by the need to evolve past/contemporary testing methodologies accordingly to such emerging systems requirements~\cite{8194898}.
The inherent complexity of these systems is amplified by the challenge of conducting testing with appropriate input data and oracles (assertions), particularly when assessing system-level functionalities~\cite{DBLP:conf/icst/Jahangirova0T21,6963470}.

A prevalent approach to testing ACPS, such as unmanned aerial vehicles (UAVs) and self-driving cars (SDCs), involves simulation environments, wherein the system under test operates within a simulated physical world~\cite{10195878,DBLP:conf/icst/KhatiriPT23,DBLP:journals/tosem/BirchlerKDPP23}.
Nevertheless, the applicability (or transferability) of traditional software testing techniques to such contexts remains unclear.
Novel and open challenges arise when dealing with the simulation-based testing of ACPSs, including the level of realism of simulations, computational costs, the complexity of simulators, and the \emph{Oracle Problem}.

Simulation-based testing research has garnered significant attention in recent years.
Particularly in exploring the challenges of test generation for simulation-based tests in the domains of UAVs and SDCs~\cite{DBLP:conf/icst/KhatiriPT23, DBLP:conf/iros/ParraO0H23, DBLP:conf/icse/BiagiolaKPR23, DBLP:conf/sbst/GambiJRZ22, DBLP:conf/sbst/PanichellaGZR21}.
This research represents a fundamental groundwork for understanding the challenges and needs inherent in simulation-based testing methodologies and testing practices for ACPSs. 
Search-based software testing techniques are a notable aspect of this research.
These techniques have proven important results in simulation-based testing, facilitating advancements in test generation and improvement~\cite{DBLP:journals/tse/FormicaFRPLM24}.
Furthermore, these search-based techniques find application beyond test generation, extending into regression testing tasks such as test minimization, selection, and prioritization \cite{DBLP:journals/tosem/BirchlerKDPP23,DBLP:conf/splc/ArrietaWSE16,ZHANG20191}.

Despite the progress made in simulation-based testing, several challenges persist.
One challenge is the \emph{Reality Gap}~\cite{DBLP:conf/icst/KhatiriPT23,afzal2021simulation,NgoBR21,RewayHWHKR20}, which refers to the disjunction between simulated environments and real-world conditions.
Additionally, issues like the \emph{Oracle Problem} and the infinite input space for simulation-based test cases pose large obstacles to the efficacy of testing~\cite{6963470}. 

In light of these challenges, this paper proposes a roadmap for future research on simulation-based testing.
Firstly, there is a pressing need to establish a clear definition and formulation of test cases tailored to simulation environments.
By delineating the characteristics and requirements of simulation-based test cases, researchers can lay the groundwork for standardized testing methodologies.
Formulating test cases for simulation environments necessitates careful consideration of various factors, including environmental dynamics, system behavior, and performance metrics.

Testing autonomous systems within simulation environments presents unique challenges.
Addressing these challenges requires innovative approaches to test case design and evaluation methodologies.
Furthermore, the paper advocates for creating and disseminating openly available benchmarks for simulation-based testing.
Establishing standards can facilitate benchmarking and comparison of different testing methodologies, fostering collaboration and knowledge sharing within the research community.
By making benchmarks openly available, researchers can streamline the testing process and promote transparency and reproducibility in experimental evaluations.

Overall, the field of simulation-based testing stands at a critical moment in time that needs further innovations and advancements.
By addressing the challenges outlined in this paper and adhering to the proposed roadmap, researchers can contribute to this expected, relevant progress towards more robust and reliable testing methodologies for autonomous systems in simulation environments.

\section{Test case formulation}\label{sec:test-case}
In this section, we provide concise software testing definitions particularly tailored for simulation-based testing of ACPSs. 
Concretley, we define a test $\theta$ as a 4-tuple:
\[ \theta=(S,E,T,O), \]
where $S$ reflects the test \emph{subject} (i.e., the system under test), $E$ the subject's \emph{environment}, $T$ the \emph{task} for the subject, and $O$ the \emph{oracle} that asserts the expected behavior of the subject. 
We describe each of these elements in detaul as follow.

\textbf{Subject.}
First, let us define the set $U$ containing all elements of our universe:
\[U=\{x \, | \, \text{x is in the universe}\}\]

The system under test is the \subject of the test case.
Formally, we define the subject as the following set:

\[S=\{x \in U \, | \,x \,\text{is part of the system under test}\}\]

In traditional software testing, we have several levels of tests.
For instance, \emph{unit testing},  which tests a piece of code within a function, or \emph{integration testing}, which tests the interaction between those units, but there are also component and system tests on higher abstraction levels.
We have different test subjects on each of those levels, i.e., different code pieces.

In the case of autonomous systems, we can test not only code and its correct execution but also its AI models, the sensor interfaces, or the interaction between physical and software components.
Furthermore, with different configuration options of those systems, we have various variants, which should be seen as different test subjects.
It is clear that defining the test subject $S$ is more complex and not as intuitive as for traditional systems, since the test subject for general ACPSs does not merely consist on its source code.

\textbf{Environment.}
Next to the test subject, we have an \emph{environment} $E$ that embeds the test subject.
We define the environment as follows:

\[ E = \{x \, | \, x \in U \setminus S \} \]

The environment covers everything except the test subject $S$.
In practice, the environment is often simplified, focusing solely on the operating system or hardware configurations.
All other aspects of the universe $U$ are omitted as they are irrelevant for the test subject.
However, in the case of simulation-based testing of ACPS, the environment usually has a higher cardinality (\ie larger) as we model the physical world with simulators.

\textbf{Task.}
Every test case has a \emph{task} $T$ that the subject has to do. Formally, we define a task $T$ as a sequence of actions.
Hence, we can write the following:

\[ T = (a_0, a_1, a_2, ..., a_n), \, n \in \mathbb{N}_0 \]

In traditional software testing, the test code (\eg unit test) sets the environment and triggers the subject to perform an action $a$ (i.e., calling the function of interest). 
In simulation-based testing, defining the test and its tasks solely through code is rarely feasible.
Simulators typically demand additional configuration files and scripts to interact with the simulation environment and describe tasks.
Furthermore, tasks involve numerous smaller actions, such as setting waypoints to define an ACPS test track.
So conceptually speaking, simulation-based tests have longer sequences defining the task than in traditional software testing, i.e., $n_{trad} < n_{sim}$.

\textbf{Oracle.}
Software engineers have expectations of how the subject has to behave.
We use the term \emph{oracle} for those expectations of the subjects within a test.
First, let us define $B$ as the set of all possible behaviors.
Hence, the oracle $O$ is a map defined as follows:

\[ O: B \times B \rightarrow \{0,1\} \]
\[ (b_{expected}, b_{actual}) \mapsto
\begin{cases}
0 & \text{if} \, b_{expected} \neq b_{actual} \\
1 & \text{if} \, b_{expected} = b_{actual}
\end{cases}
\]

In the case of a unit test of a function \texttt{sum} that accepts two arguments \texttt{a} and \texttt{b}, we expect to get the sum of these arguments.
However, in simulation-based testing, the oracle does not check if a function returns the correct value but assesses how the subject behaves in the simulation environment.
Hence, the concept of an oracle in simulation-based testing is more complex as we have to model behaviors with a sufficient abstraction to assert them with the actual observed ones~\cite{vr-paper}.

\section{Challenges}\label{sec:challenges}
Applying concepts from Section~\ref{sec:test-case} to ACPSs poses new automated testing challenges due to their complex, diverse simulation and real-world environments, as well as the test subject's complexity.

\textbf{Defining the testing task and the oracle.}
In software testing, engineers execute task actions $T$ with specific argument values and observe the behavior $b_{actual}$.
Well-defined metrics compare the actual behavior with the expected behavior $b_{expected}$.
For instance, checking the outcome of a function on equality with the expected value.
Verifying the software system behavior $b_{actual}$ against the correct behavior $b_{expected}$ is only partially automatable, i.e., automatically defining the map $O$.
This challenge is known as the \emph{Oracle Problem}.
In ACPS simulation-based testing, defining task $T$ and asserting with oracle $O$ is challenging.
Unlike traditional testing, simulation-based testing offers more freedom in defining $T$ and $O$.
However, exploring the testing space of the physical world in simulation is computationally expensive and often not cost-effective.
Moreover, simulations may lower the required computing power and the realism level of simulating the physical world.
This represents a multifaced problem for simulation-based testing:
On one hand, low simulation costs may lead to unrealistic actual behavior $b_{actual}$.
On the other hand, expensive simulation may lead to more realistic behavior $b_{actual}$.
Thus, addressing the \emph{Oracle Problem} cost-effectively in simulation-based testing is complex and requires addressing totally new research challenges.


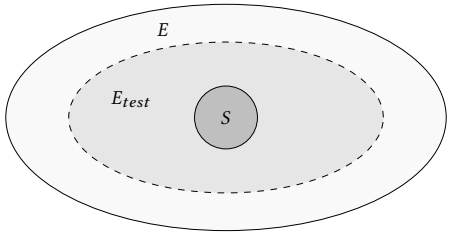
\begin{figure}
    \centering
    \resizebox{0.7\linewidth}{!}{\begin{tikzpicture}
\filldraw[fill=gray!5] (4,2) ellipse (3.5 and 1.8) node at (3,3.4) {$E$};
\filldraw[fill=gray!20!, dashed] (4,2) ellipse (2.5 and 1.2) node at (2.5,2.3) {$E_{test}$};
\filldraw[fill=gray!50] (4,2) circle (0.5) node {$S$};
\end{tikzpicture}        }
    \caption{The test subject $S$ is embedded in an environment $E$, from which we abstract many aspects away und only use a fraction as $E_{test}$.}
    \label{fig:enter-label}
\end{figure}

\textbf{Defining the environment.}
In traditional software testing, engineers create test cases using the same programming language as the subject.
Technologies such as \emph{Docker} ensure a similar testing environment to production: Real-world factors can be typically ignored/overlooked in traditional testing, so engineers focus on a subset $E_{test}$, including \emph{Docker}, OS, and hardware. 

Simulation-based testing adds complexity by the need to model the entire physical world.
When testing ACPS, software engineers must consider the physical world as the execution environment $E$ of an ACPS, which is the subject $S$.
With this, two major challenges occur:
\begin{enumerate}
    \item What aspects of the environment $E$, i.e., the physical world, can be abstracted away so that we have the testing environment $E_{test}$?
    \item How do we simulate $E_{test}$ as realistic as possible to accurately verify the subject's behavior $b_{actual}$?
\end{enumerate}

Software engineers and computer scientists work with abstractions to focus on specific aspects of their work.
In software testing, they abstract away environmental complexity to validate system behavior reliably.
Once the test environment $T_{test}$ is defined, a simulator must adequately replicate it.
However, due to the nature of potentially inaccurate/simplified simulation environments, the behavior of the subject $S$ in simulation environments $E_{test}^{sim}$ may not always reflect the behavior in the real-world environment $E_{test}^{real}$;
this leads to the \emph{Reality Gap} problem.

\textbf{Reality gap.}\label{sec:reality-gap}
In simulation-based testing, the \emph{Reality Gap}\cite{DBLP:conf/icst/KhatiriPT23,afzal2021simulation,NgoBR21} poses a critical concern.
Simulated contexts often fail to faithfully mirror real-world situations due to simplifications necessary for computational feasibility.
This trade-off between accuracy and computational time determines the extent to which simulations reflect reality\cite{collins2020traversing}.
Robotics simulations particularly struggle with accurately replicating phenomena such as actuators (e.g., torque characteristics, gear backlash), sensors (e.g., noise, latency), and rendered images (e.g., reflections, refraction, textures).
Depending on the context, the reality gap can be quantified, \eg measuring the difference a trajectory of a subject $S$ between a physical test environment $E_{test}^{real}$ and a simulated environment $E_{test}^{sim}$
This disparity between reality and simulation is known as the reality gap~\cite{collins2020traversing}.

The reality gap has been an open research problem in robotics for years now.
With the boost of \emph{Evolutionary Robotics} and the application of reinforcement learning in designing robotic control systems in recent decades, practitioners rely more and more on simulations to evaluate their designs,\ie test subjects~\cite{salvato2021crossing,hofer2021sim2real}.
More specifically, a test subject's \emph{fitness} (\eg algorithm, trained model) is calculated based on its performance in simulation for reaching the robot goals.    
However, transferring robot skills acquired in a simulated environment to a physical setting, widely referred to as \emph{Sim2Real transfer}, remains yet an open challenge~\cite{dimitropoulos2022brief}.

\textbf{Lack of Benchmarks.} 
Because of the high complexity of autonomous systems and their testing infrastructure, benchmark artifacts (e.g., simulation logs, and implementations of test subjects) are rarely openly available for research.
Furthermore, simulation-based testing is costly and hardly accessible to all researchers.
Hence, those researchers rely on openly available datasets and benchmarks.
A few recent examples in the domain of self-driving cars are SensoDat~\cite{sensodat} and DeepScenario~\cite{DBLP:conf/msr/LuYA23};
they provide datasets of driving scenarios and logged sensor data.
However, the development and testing of ACPS rely on larger comprehensive datasets to train and evaluate the various AI technologies that are part of the ACPS.

\textbf{Need of cost-effective solutions.}
As simulation-based testing for ACPS is inherently costly and non-sustainable, we require strategies to address this issue.
Traditional software testing practices like agile culture, test-driven development (TDD), DevOps methodologies, and regression testing offer quick feedback loops for developers.
Yet, adapting these techniques for simulation-based tests is uncertain.
Another challenge is making ACPS development more agile by integrating fast feedback loops with system performance in a DevOps cycle.
Undoubtedly, the research aims to make simulation-based testing more cost-effective.

\section{Automation needs \& future work}
This section discusses future research and automation needs for ACPSs, focusing on incorporating simulation-based testing into development processes and expanding the concept of test quality.

\textbf{Development and testing practices \& paradigms.} 
Agile software development fosters iterative development and rapid feedback, aiding in adapting to new requirements. Test-Driven Development (TDD) ensures systematic testing of new requirements, where test cases precede feature implementation. Despite TDD's benefits, its applicability to simulation-based testing of ACPSs remains uncertain. Ideally, TDD bridges the reality gap, ensuring ACPS behavior aligns with requirements. While numerous test cases are generated via TDD, not all need execution for each system change, following the principle of \emph{regression testing}.
With regression testing, we run for a change to the system only relevant test cases that assess the new change's behavior and verify the existing functionality's correctness.
To do so, techniques such as test minimization, selection, and prioritization are applied~\cite{DBLP:journals/stvr/YooH12}.
Applying regression testing techniques to simulation-based methods is challenging due to the requirement for computable metrics and features. Further research, as demonstrated by~\cite{DBLP:journals/tosem/BirchlerKDPP23,DBLP:journals/ese/BirchlerKBGP23}, is needed to develop such metrics and features for simulations.

Effective test cases are essential for identifying bugs and ensuring the robustness and behavior of the system. In traditional software testing, \emph{Mutation Testing} injects artificial bugs into the test subject $S$ to observe which test cases catch them~\cite{DBLP:journals/tse/Howden82}. However, applying mutation testing to ACPS in simulation is challenging due to the complexity of defining oracles in the simulated physical environment. Integrating these paradigms into ACPS development and testing can help in providing faster feedback loops for developers. Complementary, future work should address scalability and integration into DevOps steps to make these paradigms more practical, with specific attention on addressing bugs specific to ACPSs~\cite{DiSorboTOSEM2023,ZAMPETTI2022111425}. 

\textbf{Representative oracle metrics.} 
Recent research has focused on generating or improving oracles~\cite{DBLP:conf/wcre/ArrietaOHSAA22,DBLP:conf/issta/JahangirovaCHT16} for specific contexts, such self-driving cars, by simulating only the road shape. However, there is still no fully automated approach to mitigate the \emph{Oracle Problem} for ACPS contexts. Human involvement is still necessary to evaluate safety and quality. For example, in simulation-based testing for ACPSs, metrics such those in~\cite{surrealist,DBLP:conf/icst/Jahangirova0T21} are used, but recent research questions their static adoption~\cite{vr-paper}. Safety assessment via metrics may not always align with human perception due to the subjective nature of safety and its relation to realism and human experience~\cite{vr-paper}. Future research must consider what truly ensures ACPS safety and how it can be measured with quantifiable metrics.

We suggest \emph{co-simulation}~\cite{10.1145/3179993} to evaluate the subject's behavior, as $E_{test}$ is only approximated with simulations.
Co-simulation involves multiple environments with varied physical behaviors, enhancing the robustness and determinism of oracles.
Thus, test cases exhibit consistent behavior across multiple executions.

\textbf{Bridging the Reality Gap.} 
%
\emph{Domain Randomization} techniques address the reality gap by exposing algorithms to diverse random simulation environments, assuming real-world variability.
This approach aims for robustness across environments and easy transferability~\cite{james2017transferring,lee2020camera}.
Others advocate combining simulated evaluation with a small amount of real-world data~\cite{bousmalis2018using,zhang2019adversarial}, typically by recalculating fitness for selected solutions in the real world and integrating their deviation into optimization processes.
Some methods optimize general physics simulators based on real-world data, updating default settings using optimal values from real-world measurements~\cite{collins2020traversing}.
Future directions may involve exploring hybrid approaches that seamlessly blend simulated and real-world data, leveraging advancements in reinforcement learning and transfer learning techniques.
Advancements in hardware capabilities could enable more sophisticated simulation environments, enhancing variability and fidelity in training scenarios.
Incorporating domain adaptation methods from machine learning could effectively bridge the gap between simulated and real-world environments.


Previous research has highlighted challenges with simulators for testing CPS, including the reality gap, engineering complexity of realistic environments, and replicating real-world bugs~\cite{afzal2021simulation,wang2021exploratory,afzal2020study}. While solutions exist for the development phase, few address testing: \textit{How can simulators be better utilized for CPS testing, given the reality gap?}
Hildebrandt \etal~\cite{elbaum2021world} propose a mixed-reality method called \emph{world-in-the-loop} simulation for UAV testing, integrating sensor data from simulated and real environments to enhance simulation realism and diversify real-world testing.
Khatiri \etal's \textit{Surrealist} approach \cite{DBLP:conf/icst/KhatiriPT23} enables realistic simulation-based UAV testing by replicating real flights and accurately reconstructing surroundings, facilitating the identification of challenging test cases resembling real-world scenarios.
They also developed Aerialist \cite{khatiri2024simulation}, the first UAV test bench and test generation platform \cite{khatiri2024sbft}, offering new research opportunities in the field.


\textbf{Sustainable testing practices for ACPS.}
We recognize the importance of dependable ACPSs, but must also consider the influence of quality assurance tools and technologies on climate. The ICT sector is projected to produce 5.5\% of global carbon emissions and consume 20\% of all electricity by 2025~\cite{andrae2015global}. With the growing adoption of data-intensive technologies and software in daily life, software energy consumption is expected to rise.
Software testing consumes many resources regarding infrastructure and tools.
For instance, tools and techniques for creating, prioritizing, or running tests in continuous integration and deployment pipelines \cite{GranoLPP21,DBLP:journals/ese/BirchlerKBGP23,DBLP:journals/tosem/BirchlerKDPP23}.
Zaidman \etal~\cite{zaidman2024inconvenient} estimated the energy impact of various software projects and found that the \emph{Elasticsearch} project was built 5025 times in 2022.
Hence, it consumed 161.5 kWh of electricity for building the project, which corresponds to 9.7\% of the yearly average household energy consumption of a European Union citizen~\cite{zaidman2024inconvenient}.

Recent work in software engineering has highlighted \emph{green coding} practices and energy patterns for source code~\cite{manotas2016empirical,georgiou2019software, rani2024energy}.
However, energy patterns for software testing remain unexplored, along with the energy impact of testing practices.
Questions arise: How frequently should source code be built? How can regression test suites be built energy-efficiently~\cite{rani2024energy}?
Furthermore, what is the awareness among ACPS developers regarding the energy consumption of their software, and what strategies can reduce energy consumption?
Future research should explore questions to aid developers in reducing the energy footprint of software and hardware components in such systems.

\section{Conclusion}

Simulation-based testing is a standard method for evaluating the safety and quality of ACPS. While it shares similarities with traditional software testing, it's inherently more complex and can exacerbate existing issues or introduce new ones, such as the \emph{Reality Gap} (Section~\ref{sec:reality-gap}).
This paper aims to establish a unified definition and framework for testing, discussing both traditional and simulation-based approaches. It also identifies areas for future research, highlighting the challenges and automation requirements to ensure the development of safer and more sustainable/reliable ACPS for our society.

\begin{acks}
We thank the Horizon 2020 (EU Commission) support for the \href{https://www.cosmos-devops.org/}{COSMOS} project, Project No. 957254-COSMOS.
\end{acks}

\bibliographystyle{acm-template/ACM-Reference-Format}
\bibliography{main}

\end{document}